\documentclass[runningheads]{llncs}
\usepackage[T1]{fontenc}
\usepackage{graphicx}
\usepackage{multirow}
\usepackage{hyperref}
\usepackage{array}
\usepackage{cite}
\usepackage{enumitem}
\usepackage{caption}
\usepackage{enumitem}
\usepackage{subcaption}
\usepackage{xcolor}
\newcommand*\samethanks[1][\value{footnote}]{\footnotemark[#1]}
\begin{document}
\title{Effectiveness of Transformer Models on IoT Security Detection in StackOverflow Discussions}
\titlerunning{Effectiveness of Transformer models on IoT Security Detection}
%
\author{Nibir Chandra Mandal \and
G. M. Shahariar\thanks{Both authors contributed equally to this research.} \and
Md. Tanvir Rouf Shawon\samethanks}  
\authorrunning{Mandal et al.}
%
\institute{Ahsanullah University of Science and Technology, Bangladesh\\
\email{ nibir338@gmail.com, sshibli745@gmail.com, shawontanvir95@gmail.com
}}

%
\maketitle              
\begin{abstract}
The Internet of Things (IoT) is an emerging concept that directly links to the billions of physical items, or ``things", that are connected to the Internet and are all gathering and exchanging information between devices and systems. However, IoT devices were not built with security in mind, which might lead to security vulnerabilities in a multi-device system. Traditionally, we investigated IoT issues by polling IoT developers and specialists. This technique, however, is not scalable since surveying all IoT developers is not feasible. Another way to look into IoT issues is to look at IoT developer discussions on major online development forums like Stack Overflow (SO). However, finding discussions that are relevant to IoT issues is challenging since they are frequently not categorized with IoT-related terms. In this paper, we present the "IoT Security Dataset", a domain-specific dataset of 7147 samples focused solely on IoT security discussions. As there are no automated tools to label these samples, we manually labeled them. We further employed multiple transformer models to automatically detect security discussions. Through rigorous investigations, we found that IoT security discussions are different and more complex than traditional security discussions. We demonstrated a considerable performance loss (up to 44\%) of transformer models on cross-domain datasets when we transferred knowledge from a general-purpose dataset  "Opiner", supporting our claim. Thus, we built a domain-specific IoT security detector with an F1-Score of 0.69. We have made the dataset public in the hope that developers would learn more about the security discussion and vendors would enhance their concerns about product security . The dataset can be found at - \url{https://anonymous.4open.science/r/IoT-Security-Dataset-8E35}

\end{abstract}
\vspace{-5mm}
\section{Introduction}
The IoT (Internet of Things) is a widespread connection of intelligent devices that are bridged through the internet and are rapidly expanding to every corner of the globe. The number of IoT devices will touch a $36.8$ billion margin by 2025, which is $107\%$ higher than in 2020 \cite{iot_num}. Because IoT devices carry a lot of information about their users, the growing number of IoT devices presents a lot of security concerns for developers. An unintentional intrusion into the device might cause significant harm to the users. As a result, today's developers are paying close attention to this scenario and attempting to identify the best solution for protecting devices from attackers. In numerous developer communities, there is a lot of discussion about the security of IoT devices. One of the communities that developers use to address security solutions is Stack Overflow (SO). The SO community is quite active and has a lot of IoT security discussions. A great number of works have been done in this research field \cite{barua}\cite{Bagherzadeh}\cite{yang}. In this work, we have tried to demonstrate a comprehensive analysis of determining whether an aspect of a discussion is an IoT security aspect or not. The purpose of this study is to assist developers in learning more about security issues in the IoT space, where there is a lot of discussion about various aspects. Furthermore, with the aid of our efforts, vendors may increase the security of their products. We have basically employed some of the popular transformer models like BERT\cite{bert}, RoBERTa\cite{roberta}, XLNet\cite{xlnet} and BERTOverflow\cite{BERTOverflow} in $3$ of our different experiments. For our initial approach, we used a benchmark dataset called Opiner \cite{opiner}, which recorded a variety of aspects of discussions such as performance, usability, security, community, and so on. The field of discussion was also diverse. Instead of a general-purpose dataset, we worked to create a domain-specific dataset focused solely on IoT security issues, which we named the "IoT Security Dataset". Following the creation of the dataset, we attempted to transfer the knowledge gained from the Opiner dataset to our own IoT dataset, but the results were unsatisfactory. Then we decided to fine-tune the pretrained tansformer models using our own dataset, and we achieved a decent result. From these experiments, we discovered that knowledge transfer is not a good technique to get a decent outcome. The best way to identify security issues in developer discussions is to conduct domain-specific experiments using domain-specific data.  
In summary, we have made the following main contributions in this paper: 
\begin{itemize}
    \item We have created a domain-specific dataset called "IoT Security Dataset" that focuses on security aspects of IoT related textual developer discussions.
    \item We further employed multiple transformer models to automatically detect security discussions. Through rigorous investigations, we found that IoT security discussions are different and more complex than traditional security discussions.
    \item We demonstrated a considerable performance loss (up to 44\%) of transformer models on cross-domain datasets when we transferred knowledge from a general-purpose dataset "Opiner", supporting our claim.
\end{itemize}

\section{Background Study}
\label{sec:BStudy}
\subsection{Transformers}
A neural network that learns factors in sequential data is referred to as a transformer model. A transformer model provide us with an word embedding which is used as the input of a fully connected neural network for further classification task. In natural language processing transformer models are used frequently as it tracks the meaning of consecutive data by tracking the relationships among them like the words in a sentence. Transformer models were first introduced by Google \cite{transformer} and are one of the most powerful models till date in the history of computer science.
We incorporated four different kinds of transformer models in our work which are RoBERTa, BERT, XLNET and BERTOverflow. Bidirectional Encoder Representations from Transformers or BERT \cite{bert} was first established by Jacob Devlin et al. which is a transformer model to represent language. It was modelled in such way that anyone can fine tune it with a supplementary layer to create new models to solve a wide range of tasks. 
Yinhan Liu et al. introduced RoBERTa \cite{roberta} as a replication study of BERT \cite{bert}. Authors showed that their model was highly trained than BERT overcoming the limitations and showed a good performance over basic BERT model.
Zhilin Yang et al. introduced XLNet \cite{xlnet} which overcomes the constraints of BERT \cite{bert} using a universal autoregressive pre-training technique. Maximum expected likelihood was taken into consideration over all arrangements of the factorization order. 
Authors showed that their model beat BERT by a huge margin on a variety of tasks like sentiment analysis, language inference or document ranking etc.
BERTOverflow \cite{BERTOverflow} by Jeniya Tabassum et al. is a kind of transformer model trained for identifying code tokens or software oriented entities inside the area of natural language sentences. Authors introduced a NER corpus of total $15,372$ sentences for the domain of computer programming. 

\subsection{Evaluation matrices}
Any kind of model requires some measurement criteria to understand its' efficiency. In this project $4$ different kinds of performance matrices were used which are \textbf{Precision}, \textbf{Recall}, \textbf{F1-Score} and \textbf{Area Under Curve (AUC)}.
The percentage of accurately predicted labels that are literally correct is calculated by a model's \textbf{precision} score. It also indicates the calibre of a model to predict the labels correctly. The model's ability to reliably discover positives among existing positives is measured by its \textbf{recall} score. We can also state it as a model's capacity to locate all relevant cases within a data set. The precision and recall score are used to generate the \textbf{F1-Score}. The F1-score is a metric in which Precision and Recall play an equal role in evaluating the model's performance. A graphical illustration of how effectively binary classifiers work is the Receiver Operating Characteristic (ROC) curve. \textbf{(AUC)} score is calculated using ROC and is a very useful tool for calculating a model's efficiency in an uneven dataset.

\subsection{Cross validation}
A reshuffling process for assess various algorithms and models on a little amount of data is known as Cross-validation. It comprises an attribute, k, denoting the quantity of sets a provided sample of data should be partitioned. That is why, the technique is widely known as k-fold cross-validation. When we give a value to the parameter k, it means we want to partition the whole data into k sets. 

\section{The IoT Security Dataset}
\label{sec:dataset}
In this section, we will present details of the IoT security 
dataset creation process. The whole process is divided into three sub\-processes: (1) Collect IoT posts (2) Extract sentences from IoT posts and (3) Label extracted sentences. Each process is discussed in the following sub\-sections. 
\vspace{-3mm}
\subsection{IoT Posts Collection}
As IoT tags do not cover all IoT related posts, we applied three steps to collect IoT related posts. 
\begin{enumerate}[label=\textbf{Step \arabic{enumi})}, wide, labelwidth=!,itemindent=!, labelindent=0pt]
\item \textbf{Download the SO dataset:} We used the September 2021 Stack Overflow (SO) data dump, which was the most recent dump accessible at the time of research. The following meta data may be found in each post: (1) \textit{text data and code samples}; (2) \textit{creation and edition times}; (3) \textit{score, favorite, and view counts}; (4) \textit{user information of the post creator}; and (5) \textit{if the post is a question, tags given by the user}. If the user who posted the question marked the response as approved, the answer is flagged as accepted. A question might include anything from one to five tags. All postings from 2008 to September 2021 are included in the SO dataset.
\item \textbf{Identify IoT Tagset:} Because not all of the posts in the SO dataset are linked to the IoT, we had to figure out which ones had IoT-related topics. We established the set of tags that may be used to label a discussion as IoT-related using the user-defined tags attached to the questions. We followed the two-step method described in \cite{gias_emp}: (1) In SO, we discovered three prominent IoT-related tags. (2) We gathered postings with those three initial tags and studied the tags that were applied to them. There are 78 tags in the final tag collection. We have discussed each step in detail below.
\begin{enumerate}[label=\textbf{(\alph{*})}, wide, labelwidth=!, labelindent=0pt]
\item \textbf{Identify initial tags:} Using the Stack Overflow (SO) search engine, the three initial tags were chosen. We began our search by looking for queries in SO that were labeled with "iot", and the SO search engine returned posts that were tagged with "iot" as well as a set of 25 additional tags that were related to these topics, such as "raspberry-pi", "arduino", "windows-10-iot-core", "python," and so on. In our initial set, we thus considered the following three tags: (a) "iot" or any tag including the term "iot", such as "windows-10-iot-core"; (b) "arduino", and (c) "raspberry-pi".
\item \textbf{Determine final tagset:} To begin, we looked for IoT-related posts using the three initial tags. Second, we retrieved all questions from the SO dataset that were tagged with one of the initial three tags. Finally, we collected all of the tags from the question set. It is possible that not all of the tags in the questions set match to IoT subjects (for example, "python"). As a result, the significance and relevance specified in \cite{gias_emp} for each tag were computed to pick filtered and finalized IoT-related tags in the questions set. If a tag's significance and relevance values exceed certain criteria, it is considered significantly relevant to the IoT. After comprehensive testing, we discovered 78 IoT-related tags. Please see \cite{gias_emp} for a more complete discussion of the tagset identification technique.
\end{enumerate}
\item \textbf{Collect IoT Posts:} All posts labeled with at least one of the selected 78 tags make up our final dataset. If a SO question is labeled with one or more tags from the final tagset, it is an IoT question. In our SO dump, we discovered a total of 40K posts based on the 78 tags.
\end{enumerate}
\vspace{-3mm}
\subsection{Sentence Level Dataset Creation}
\vspace{-1mm}
As security discussions can be buried inside a post, So we followed sentence-level analysis to get a better understanding of the security discussion. We got the sentences as follows:
As our main focus is textual analysis, we avoided security discussion inside code snippets and urls. We used the BeautifulSoup library to extract all that extra information such as titles, codes, and urls. We also used the NLTK sentence tokenizer to get all the sentences from those 40K posts. Finally, we found around 200K sentences.

\subsection{Benchmark Dataset Creation}As there are not any automated tools to label this large data set, we chose to label them manually. Due to the data size and sparsity, we labeled a statistically significant portion of the data. For our experiment, we randomly selected 7147 (99\% confidence with a confidence interval of 1.5) sentences and labeled them. Without knowing each other, two developers labeled this data first. In addition, a third developer was introduced to resolve the conflict. We labeled each sentence by majority voting. For example, if both developers agreed, we labeled them with the same label they agreed on. Otherwise, we took into account the third developer’s decision. All these developers have industrial experience of more than 2 years and have enough knowledge of the domain. The summary of our dataset can be seen here in Table~\ref{tab:iot_opiner_distribution}.

\vspace{-2em}
\begin{table}[htbp]
\centering
\caption{Security Data Distribution for Opiner and IoT security datasets.}
\label{tab:iot_opiner_distribution}
\begin{tabular}{|c|c|c|c|c|}
\hline
\textbf{ Dataset } & \textbf{ Size } & \textbf{ Security } & \textbf{ Kappa } & \textbf{\begin{tabular}[c]{@{}c@{}} Agreement \\ Score \end{tabular}} \\ \hline
Opiner           &  4522          & 163 (3.6\%)       & -              & -                                                                   \\ \hline
IoT              &  7147          & 250 (3.5\%)       & 0.92           & 98.5\%                                                              \\ \hline
\end{tabular}
\end{table}
\vspace{-7mm}

\section{Opiner Dataset}
Opiner datatset was created by Uddin et al. \cite{gias_mining} to study developers opinion in SO for different APIs. The dataset includes developer discussions in form of text and related aspects of that discussions. Each opinion may have multiple aspects such as "Performance", "Usability", etc. One of those aspects is "Security". In this research, we only took consideration of this aspect. We discarded other aspects and labeled all security aspect related samples as 1 and other samples as 0. We found a total of 163 samples are security related.

\section{Methodology}
\label{sec:proposed_methodology}
The proposed system for aspect categorization is presented in this section. The key phases of the proposed method for binary aspect classification (\textit{considering usability aspect as an example}) are summarized in Figure~\ref{classifier} and are further detailed below.
\begin{enumerate}[label=\textbf{Step \arabic{enumi})}, wide, labelwidth=!, labelindent=0pt]
\item \textbf{Input Sentence}: Each raw sentence from the dataset is presented to the proposed model one by one for further processing. Before that each sentence goes through some pre-processing steps: all urls and codes are removed.
\item \textbf{Tokenization}: Each processed sentence is tokenized using the \textit{BERT Tokenizer}. Each tokenized sentence gets a length of 100 tokens and zero padded when required. In the event of length more than 100, we cut off after 100. The output of this step is a tokenized sentence of size 100.
\item \textbf{Embedding}: For word embedding, BERT \cite{bert} is used to turn each token in a sentence into a numeric value representation. Each token is embedded by 768 real values via BERT. The input to this step is a tokenized sentence of size 100 and output of this step is an embedded sentence of size 100*768.
\item \textbf{Pooling}: To reduce the dimension of the feature map (100*768) for each tokenized sentence in step 3, max pooling is used which provides a real valued vector representation of size 768 per sentence.
\vspace{-1em}
\begin{figure}
\includegraphics[width=\textwidth]{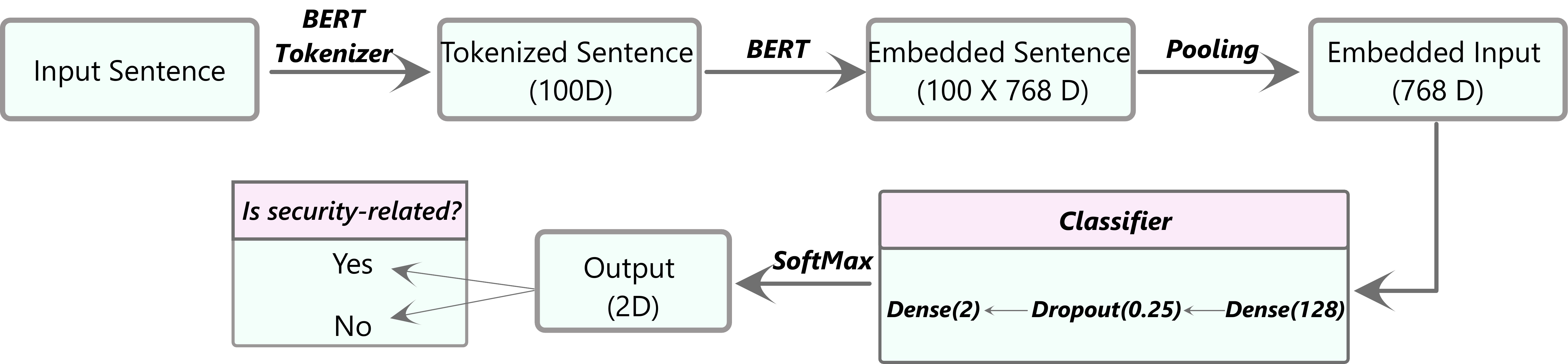}
\caption{Aspects Classification Process} \label{classifier}
\end{figure}
\vspace{-5mm}
\item \textbf{Classification:} Aspect classification is accomplished by the use of transfer learning \cite{raffel}. To classify security and non security aspects, a basic neural network with two dense layers (with 128 and 2 neurons) is utilized to fine-tune the four pretrained transformer models one by one such as RoBERTa \cite{roberta}, BERT \cite{bert}, DistilBERT \cite{distilbert}, and XLNET \cite{xlnet}. The pre\-trained weights were initialized using \textit{glorot\_uniform} in the simple neural network. A dropout rate of 0.25 is employed between the dense layers. Finally, we apply a `Softmax' activation to get the prediction.
\end{enumerate}
\vspace{-5mm}
\section{Experimental Results}
\label{sec:experiment}
\subsection{Experiments}

\begin{figure}[!ht]
		\centering
		\begin{subfigure}[b]{.49\columnwidth}
		    \centering
			\includegraphics[width=1\linewidth]{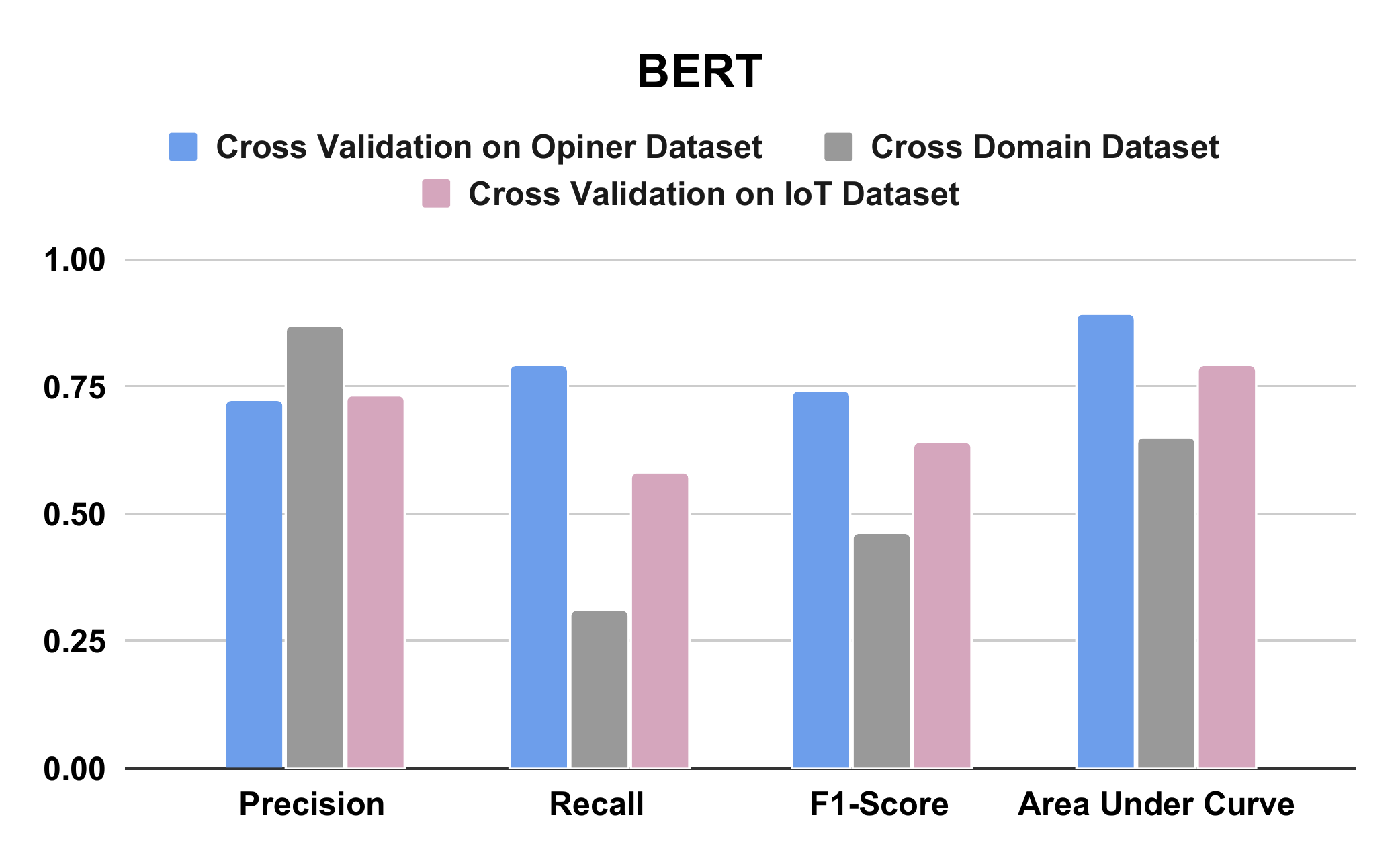}
		\end{subfigure}
		\begin{subfigure}[b]{.49\columnwidth}
		    \centering
			\includegraphics[width=1\linewidth]{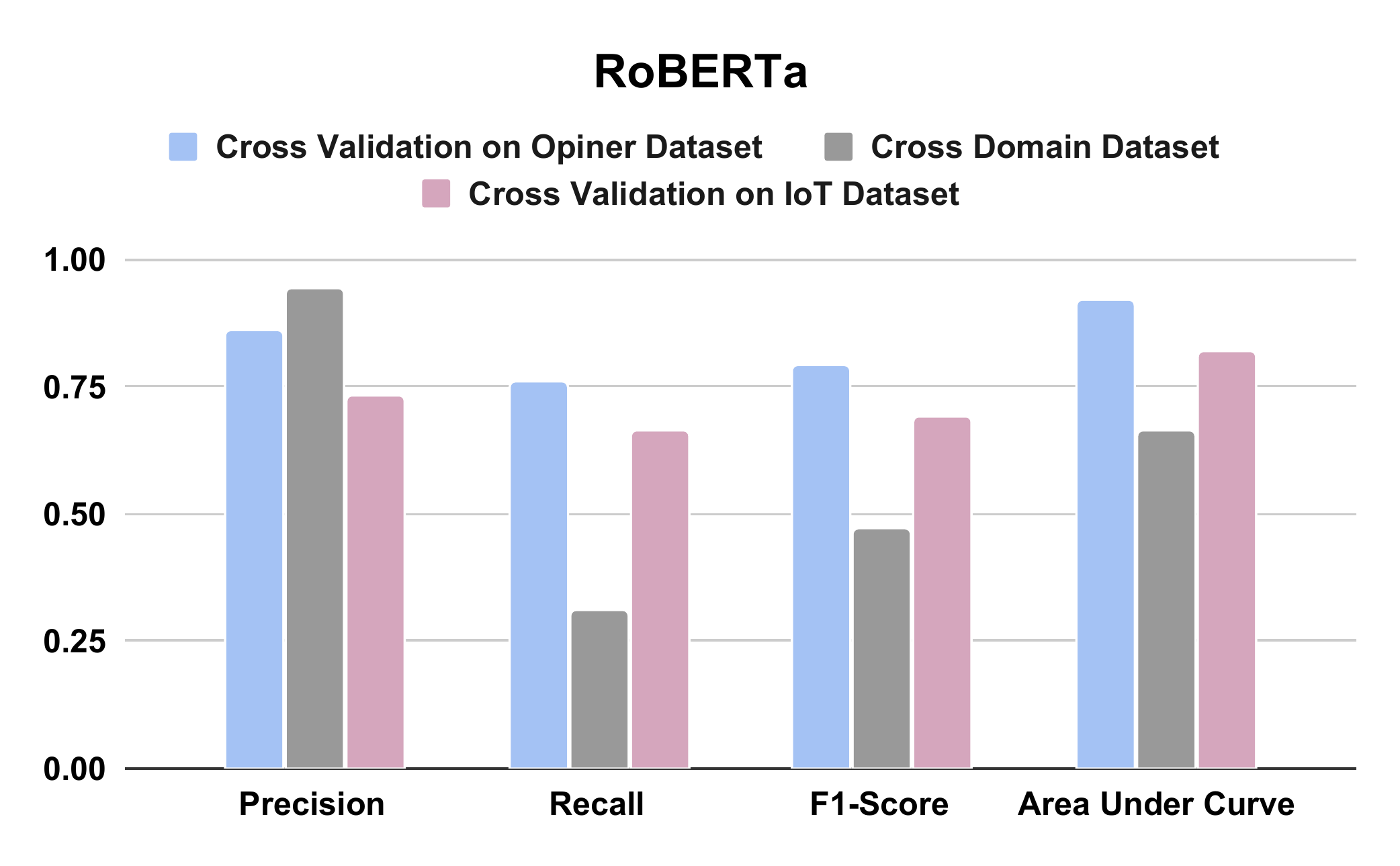}
		\end{subfigure}
		\begin{subfigure}[b]{.49\columnwidth}
		    \centering
			\includegraphics[width=1\linewidth]{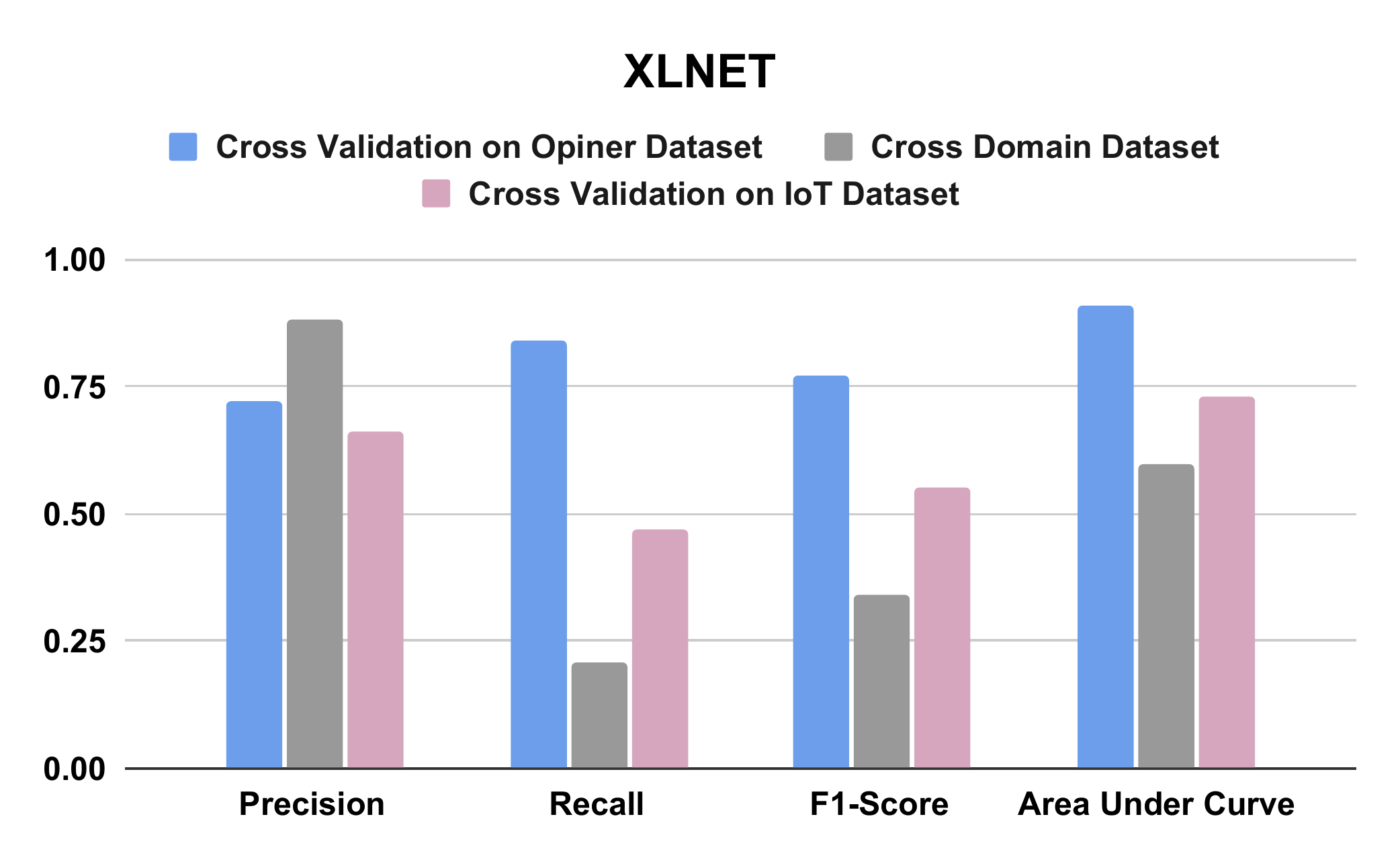}
		\end{subfigure}
		\begin{subfigure}[b]{.49\columnwidth}
		    \centering
			\includegraphics[width=1\linewidth]{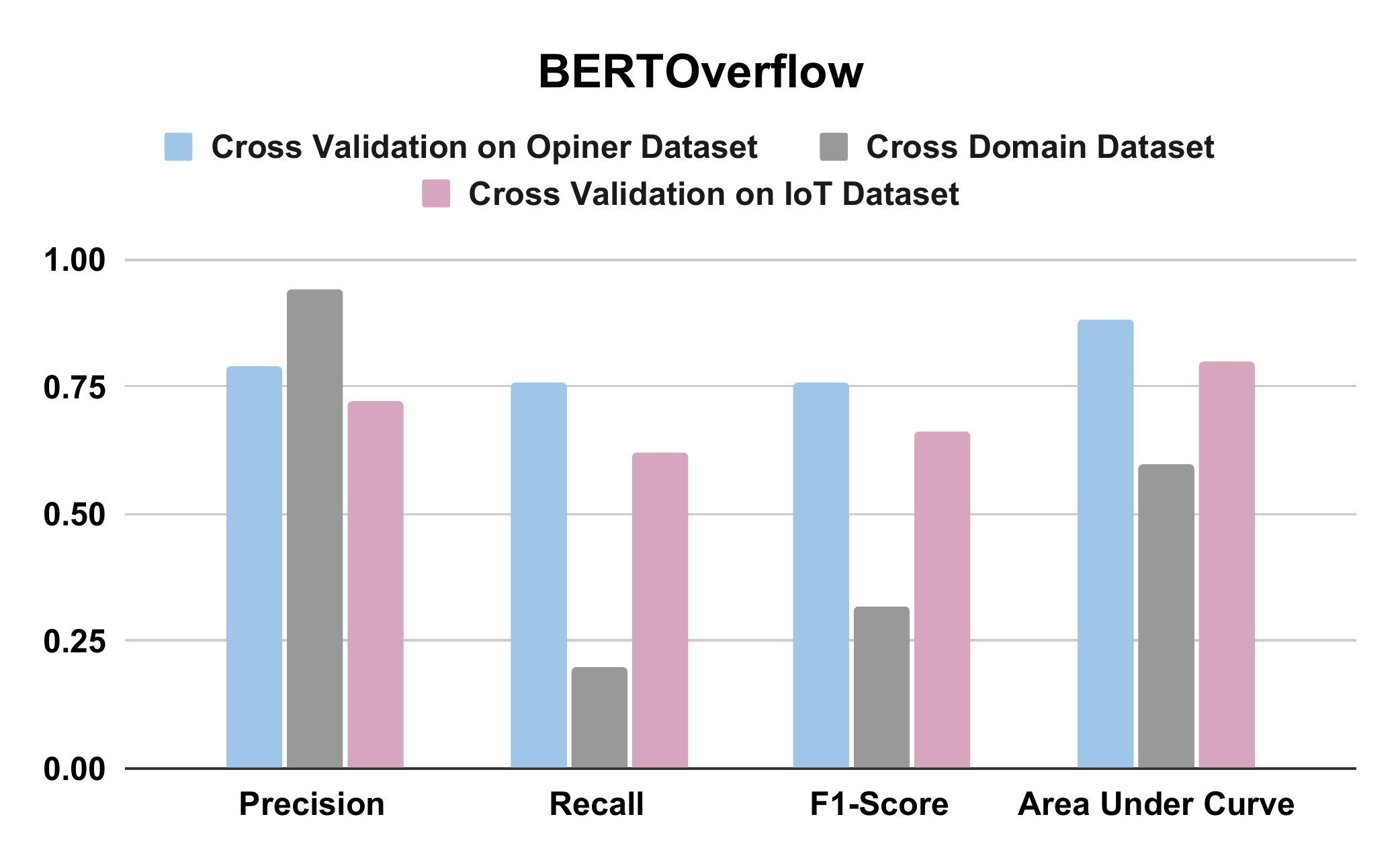}
		\end{subfigure}
		
		\setlength{\belowcaptionskip}{-15pt}
		\caption{A comparison of performance among all the $3$ experiments for BERT (\textit{top left}), RoBERTa (\textit{top right}), XLNET (\textit{lower left}) and BERTOverflow (\textit{lower right}) models. }
		\label{fig:comparison}
	\end{figure}

At first, we found that Uddin el.’s Logistic Regression (Logits) model \cite{gias_mining} outperformed other rule-based models. Thus, we took this Logits as our benchmark model. In recent times, deep models, specifically pretrained models, have shown a huge improvement over both shallow machine learning models such as SVM and Logistics Regression and classic deep models such as Long Short Term Memory (LSTM), Convolutional Neural Network (CNN), and Bidirectional Long Short Term Memory (Bi-LSTM). Therefore, we decided to use the Transformer models. We applied these already trained models on Opiner dataset to test the IoT dataset which we have prepared for this study to check whether a general purpose dataset can give a good generalization performance for a domain specific dataset. We did not get a satisfactory result from this experiment as the generalization is not a good idea. Finally, we added domain-specific IoT security data during the tuning of the pre-trained models and found acceptable results from the experiment. The performances of all the models on opiner dataset are shown in Table \ref{tab_performance}. \textbf{The result of logistic regression is taken from the experiments \cite{opiner} of Gias Uddin et al on the benchmark Opiner dataset.} A comparison of performances among all the $3$ experiments on Cross-validation on Opiner dataset, Cross-Domain dataset (Trained on Opiner and tested on IoT dataset) and Cross-validation on IoT Dataset for all the transformer models we used are shown in figure \ref{fig:comparison}.
\vspace{-1.5em}
\begin{table}[]
\caption{Performance of $4$ different transformer model on \textit{Opiner} dataset }
\label{tab_performance}
\centering
\begin{tabular}{|c|lllll|}

\hline
\textbf{Type} &\multicolumn{1}{|c|}{\textbf{Model}}        & \multicolumn{1}{c|}{\textbf{Precision}} & \multicolumn{1}{c|}{\textbf{Recall}} & \multicolumn{1}{c|}{\textbf{F1-Score}} & \multicolumn{1}{c|}{\textbf{AUC}} \\ \hline
\multirow{4}{*}{\textbf{Proposed Models}  }&\multicolumn{1}{|c|}{BERT}         & \multicolumn{1}{l|}{0.72}       & \multicolumn{1}{l|}{0.79}   & \multicolumn{1}{l|}{0.74}     & 0.89                          \\ 
&\multicolumn{1}{|c|}{RoBERTa}      & \multicolumn{1}{l|}{\textbf{0.86}}       & \multicolumn{1}{l|}{0.76}   & \multicolumn{1}{l|}{\textbf{0.79}}     & \textbf{0.92}                          \\ 
&\multicolumn{1}{|c|}{XLNet}        & \multicolumn{1}{l|}{0.72}       & \multicolumn{1}{l|}{\textbf{0.84}}   & \multicolumn{1}{l|}{0.77}     & 0.91                          \\ 
&\multicolumn{1}{|c|}{BERTOverflow} & \multicolumn{1}{l|}{0.79}       & \multicolumn{1}{l|}{0.76}   & \multicolumn{1}{l|}{0.76}     & 0.88                          \\  \hline
\textbf{Previous Work} &\multicolumn{1}{|c|}{Logits \cite{gias_mining}} & \multicolumn{1}{l|}{0.77}       & \multicolumn{1}{l|}{0.57}   & \multicolumn{1}{l|}{0.60}     & 0.69                          \\ \hline

\end{tabular}
\end{table}
\vspace{-10mm}

\subsection{Result Analysis}
\label{sec:result_analysis}
First, We experimented with three generic transformers, BERT, RoBERTa, and XLNet, as well as one domain-specific model, BERTOverflow. As these robust pretrained models can capture underlying implicit contexts, the performance of these models significantly improved over shallow models, although the size of the dataset is comparatively low. We found that the lowest performing transformer model has 23\% better F1-Score than the baseline models.

\begin{figure}[!htp]
		\centering
			\includegraphics[width=.7\linewidth]{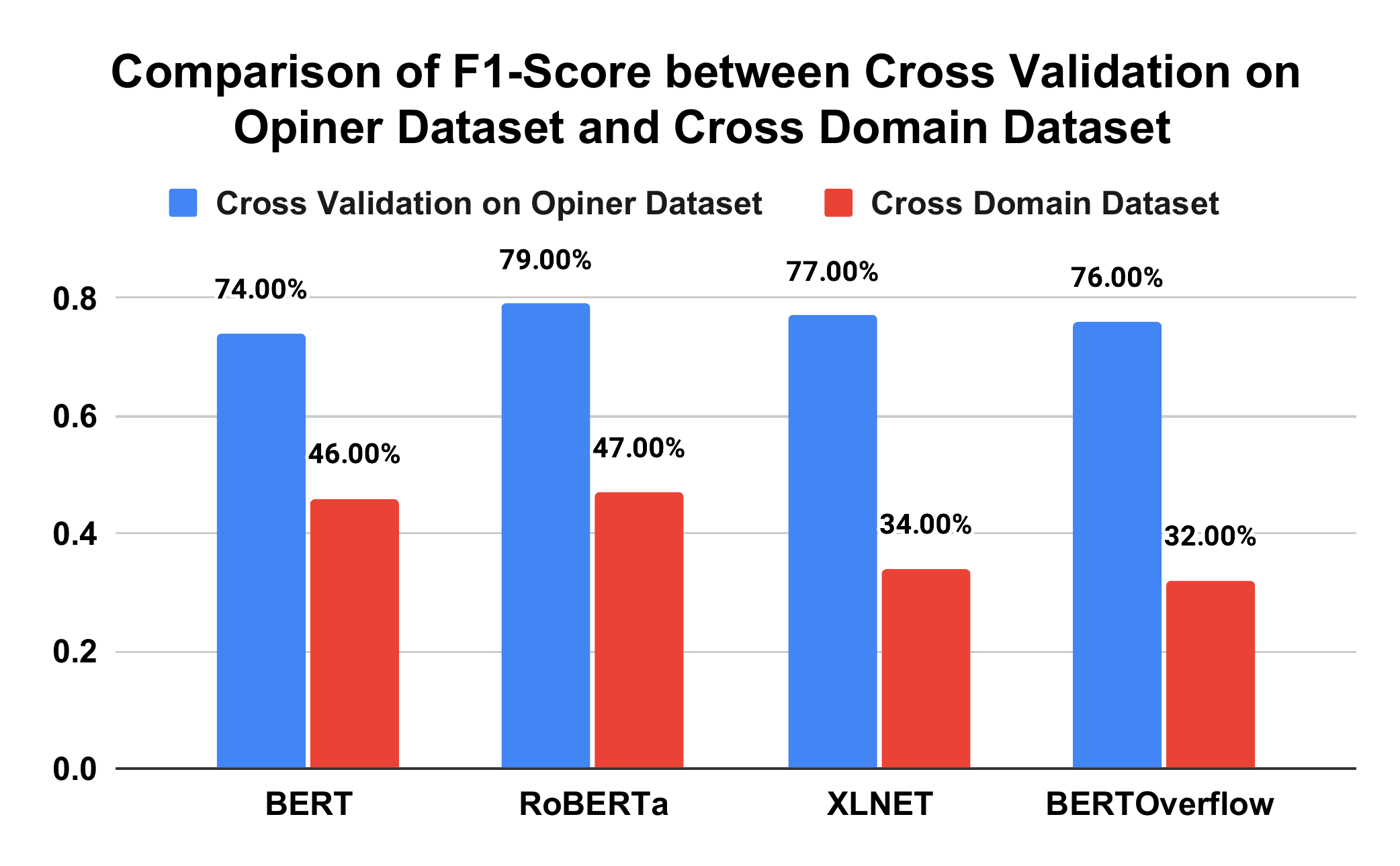}
		\setlength{\belowcaptionskip}{-15pt}
		\caption{A comparison of \textbf{F1-Score} between the experiments on Cross-validation on Opiner dataset and Cross-Domain dataset}
		\label{fig:comparison_f1}
		
	\end{figure}

\noindent 
We found that RoBERTa model has the best performance in terms of F1-Score among other transformers. 
Despite having a low recall, RoBERTa achieves the highest F1-Score of 0.79, followed by XLNet (0.77), BERTOverflow (0.76), and BERT (0.74). This indicates that RoBERTa is the most precise model, but it has lower discoverability. Thus, the generic purpose transformer RoBERTa outperforms both the domain-specific BERTOverflow and other generic purpose models. This happens for two reasons. \textit{\textbf{First,}} RoBERTa is the most optimized transformer model. Due to this optimization, RoBERTa gets an extra edge to outperform its base model, BERT, for our task.
 \textit{\textbf{Second}}, security discussions in the Opiner dataset are more generic than domain-specific. Although the traditional rule-based approach fails to guess the security discussions as correctly as deep models like the transformer model, SO domain knowledge makes the prediction more intriguing.
XLNet has enough insights to identify security discussions but it suffers from precision.
As example, we found the following sentences in Opiner dataset: \textit{`For instance, messages sent by the client, may be digitally signed by the applet, with the keys being managed on per-user USB drives (containing the private keys).'} It is clearly visible that the word `signed' is a pivotal security-related word in this sentence. As our SO domain-knowledge specific model BERTOverflow is trained on both programming and non-programming data, the model finds ambiguity in such a scenario and thus makes erroneous predictions. However, RoBERTa handles the scenario in a better way.\\
\noindent We found that transformers improved the performance of baseline models by a long way for security aspect detection. This motivated us to dive deeper into security aspects.
However, we found that the performance significantly dropped when the transformers trained on Opiner security data were tested on the IoT dataset. We investigated this drop in performance and found that security discussions in the IoT dataset are more implicit compared to the generic security discussion in Opiner. Moreover, the domain knowledge that we need to discern those security discussions is missing in the generic Opiner dataset. For instance, the following sentence includes the IoT domain specific keyword RFID which is an identification key: `Is there any way to restart the screen after using the RFID reader?'. As such discussions are only available in IoT discussions, transformer models that are created and trained on only for general purposes fail to perceive the security context. As a result, the model underperforms in the IoT domain. The performance drop in F1-Score for all the transformers can be seen in Figure \ref{fig:comparison_f1}. We can clearly see from the charts that the performance of \textbf{BERT, RoBERTa, XLNET and BERTOverflow} has dropped by \textbf{28\%, 32\%, 43\% and 44\%} respectively after testing our own IoT dataset with the transformers trained on Opiner dataset.\\
\noindent As our previous experiment result analysis denotes that the model is missing IoT contextual insights, we included IoT discussions in our dataset to check how these discussions influence the performance. We observed that adding domain-specific data during the tuning of the pretrained model resulted in more robust performance. We found that the performance improved over $50\%$ in terms of F1-score. Like our previous findings, the RoBERTa model has the best F1-Score of 0.69, followed by BERTOverflow (0.66), BERT (0.64), and XLNet (0.54). In addition, RoBERTa also shows the best precision of 0.73 and recall of 0.66. This again shows that RoBERTa is the most effective model for security aspect detection in any domain.
\vspace{0mm}
We compared this result with the results of previous experiments. We have showed bar-charts in Figure \ref{fig:comparison} for all these four models. We found that all models have lower precision than previous experimental results. However, the models had improved their recall compared to experiment 2 but still lagged behind experiment 1. As a result, the models have a higher F1-Score than experiment 2 but lower than experiment 1. For example, RoBERTa has a 46\% better F1-Score than experiment 2, but 13\% lower than experiment 1. Based on this observation, we came to two conclusions regarding the detection of IoT security aspects. \textit{\textbf{First, IoT security is different from the general type of security we discuss more often.}} We conclude this based on the evidence that the same transformers that can detect security aspects reliably (Experiment-1 results) fail to perform similarly in the IoT domain (Experiment-2 results), but the performance improves while domain knowledge is added to the model (Experiment-3 results). \textit{\textbf{Second, IoT security aspects are more complex, sparse, and implicit than normal security aspects.}} The basis of this conclusion is the evidence that even if the transformer models are adopted to IoT domain knowledge, the models fail to perform as well as they do for general security aspect detection (Experiment-3 results). Thus, we believe by incorporating more IoT knowledge (i.e., increasing training samples) the performance of the models can be further improved, which we left as our future work.

\section{Related Work}
\label{sec:related_work}
Several recent papers have used SO posts to investigate various aspects of software development using topic modeling, such as what developers are talking about in general \cite{barua}, or about a specific problem like big data \cite{Khatchadourian}, concurrency \cite{Bagherzadeh}, security \cite{yang}\cite{gias_ist}, and trends in block chain reference architectures \cite{wan}. All of these projects demonstrate that interest in the Internet of Things is growing, and conversations about it are becoming more common in online developer communities like Stack Overflow (SO). Analyzing the presence, popularity, and complexity of certain IoT issues may be gained through understanding these discussions. Uddin et al. \cite{gias_emp} investigated at nearly 53,000 IoT-related posts on SO and used topic modeling \cite{blei} to figure out what people were talking about. On SO, Aly et al. \cite{aly} addressed at questions of IoT and Industry 4.0. They utilized topic modeling to identify the themes mentioned in the investigated questions, similar to the work given by Uddin et al. \cite{gias_emp}. Their study concentrated on the industrial issues of IoT technology, whereas Uddin et al. \cite{gias_emp} intended to learn about the practical difficulties that developers experience when building real IoT systems. Recent studies \cite{kavaler, parnin} explored the connection between API usage and Stack Overflow discussions. Both research discovered a relationship between API class use and the number of Stack Overflow questions answered. But Gias Uddin \cite{gias_mining} utilized their constructed benchmark dataset named "OPINER" \cite{gias_mining} to carry out the study and noticed that developers frequently provided opinions about vastly different API aspects in those discussions which was the first step towards filling the gap of investigating the susceptibility and influence of sentiments and API aspects in the API reviews of online forum discussions.
Uddin and Khomh \cite{gias_mining} introduced OPINER, a method for mining API-related opinions and providing users with a rapid summary of the benefits and drawbacks of APIs when deciding which API to employ to implement a certain feature. Uddin and Khomh \cite{gias_summary} used an SVM-based aspect classifier and a modified Sentiment Orientation algorithm \cite{liu} to comply with API opinion mining. Based on the positive and negative results emphasized in earlier attempts to automatically mine API opinions, as well as the seminal work in this field by Uddin and Khomh \cite{gias_summary}, Lin et al. \cite{lin_bin} introduced a new approach called Pattern-based Opinion MinEr (POME), which utilizes linguistic patterns preserved in Stack Overflow sentences referring to APIs to classify whether a sentence is referring to a specific API aspect (functional, community, performance, usability, documentation, compatibility or reliability), and has a positive or negative polarity.\\
\vspace{-0.5cm}
\section{Conclusion and Future works}
\label{sec:concusion}
We found identifying security aspects in IoT-related discussions to be a difficult task since domain-specific datasets on security-related discussions are not commonly available. In our study, we attempted to create a one-of-a-kind dataset in this respect, and we presented a brief comparison between our IoT security dataset experiments and the benchmark dataset on aspect identification. We have come to the conclusion that generalization is not really the best method for identifying security discussions on sites like StackOverflow. Domain-specific knowledge transfer via various transformer models might be a superior alternative to security aspect detection. In the future, we can incorporate other transfer learning methods to improve our performance. The results we obtained are not quite satisfactory as our dataset is not very large. Increasing the number of samples in the dataset is another effort we may undertake in the future to enhance the outcome. 
%
%
%
%

\end{document}